\documentclass[prl,twocolumn,superscriptaddress]{revtex4-2}
\usepackage{amssymb}
\usepackage{amsfonts}
\usepackage{mathrsfs}
\usepackage{graphics,graphicx,epsfig,amsmath,amsthm}
\usepackage{floatrow}
\usepackage{caption}
\usepackage[labelformat=simple]{subfig}

\usepackage{upgreek}
\usepackage{bm}
\usepackage{bbm}
\usepackage{multirow}
\usepackage{array}
\usepackage{color}
\usepackage{cases}
\usepackage{booktabs }
\usepackage[usenames,dvipsnames]{xcolor}

\usepackage{xcolor}

\usepackage{float}
\usepackage[colorlinks=true,linkcolor=black,citecolor=blue,urlcolor=blue]{hyperref}
\usepackage[doipre={doi:~}]{uri}
\makeatletter
\let\saved@includegraphics\includegraphics
\AtBeginDocument{\let\includegraphics\saved@includegraphics}
\renewenvironment*{figure}{\@float{figure}}{@float}
\makeatother
\begin{document}

\title{Experimental test of the Crooks fluctuation theorem in a single nuclear spin}
\author{Wei Cheng}
\affiliation{CAS Key Laboratory of Microscale Magnetic Resonance and School of Physical Sciences, University of Science and Technology of China, Hefei 230026, China}
\affiliation{CAS Center for Excellence in Quantum Information and Quantum Physics, University of Science and Technology of China, Hefei 230026, China}
\author{Wenquan Liu}
\affiliation{School of Science, Beijing University of Posts and Telecommunications, Beijing, 100876, China}
\author{Zhibo Niu}
\affiliation{CAS Key Laboratory of Microscale Magnetic Resonance and School of Physical Sciences, University of Science and Technology of China, Hefei 230026, China}
\affiliation{CAS Center for Excellence in Quantum Information and Quantum Physics, University of Science and Technology of China, Hefei 230026, China}
\author{Chang-Kui Duan}
\affiliation{CAS Key Laboratory of Microscale Magnetic Resonance and School of Physical Sciences, University of Science and Technology of China, Hefei 230026, China}
\affiliation{CAS Center for Excellence in Quantum Information and Quantum Physics, University of Science and Technology of China, Hefei 230026, China}
\affiliation{Hefei National Laboratory, University of Science and Technology of China, Hefei 230088, China}
\author{Xing Rong}
\email{xrong@ustc.edu.cn}
\affiliation{CAS Key Laboratory of Microscale Magnetic Resonance and School of Physical Sciences, University of Science and Technology of China, Hefei 230026, China}
\affiliation{CAS Center for Excellence in Quantum Information and Quantum Physics, University of Science and Technology of China, Hefei 230026, China}
\affiliation{Hefei National Laboratory, University of Science and Technology of China, Hefei 230088, China}
\author{Jiangfeng Du}
\email{djf@ustc.edu.cn}
\affiliation{CAS Key Laboratory of Microscale Magnetic Resonance and School of Physical Sciences, University of Science and Technology of China, Hefei 230026, China}
\affiliation{CAS Center for Excellence in Quantum Information and Quantum Physics, University of Science and Technology of China, Hefei 230026, China}
\affiliation{Hefei National Laboratory, University of Science and Technology of China, Hefei 230088, China}
\affiliation{School of Physics, Zhejiang University, Hangzhou 310027, China}

\begin{abstract}
We experimentally test the Crooks fluctuation theorem in a quantum spin system. 
Our results show that the Crooks fluctuation theorem is valid for different speeds of the nonequilibrium processes and under various effective temperatures.
Work is not an observable in quantum systems, which makes tests of quantum thermodynamic theorems challenging.
In this work, we developed high-fidelity single-shot readouts of a single nuclear spin in diamond and implemented the two-point work measurement protocol, enabling a direct experimental test of the Crooks fluctuation theorem.
Our results provide a quantum insight into fluctuations and the methods we developed can be utilized to study other quantum thermodynamic theorems.
\end{abstract}
\maketitle

\section{I. Introduction}
Fluctuations become prominent when the study of thermodynamics shifts from the macroscopic to the microscopic scale.
These fluctuations can be comparable with ensemble averages of corresponding thermodynamic quantities~\cite{Baumer2018} and are not mere background noises~\cite{Jarzynski2011}.
Investigation of these fluctuations has led to the discovery of various fluctuation theorems~\cite{Sevick2008,Esposito2009,Campisi2011,Funo2018,Landi2021}.
One important example is the Crooks fluctuation theorem (CFT)~\cite{Crooks1999}, which reads as $P^F(W)/P^R(-W) = e^{\beta(W-\Delta F)}$, with $\beta$ denoting the inverse temperature.
The CFT relates the probability $P^F(W)$ of performing some work $W$ during a forward process to the probability $P^R(-W)$ of extracting the same amount of work during the time-reversed process via the free-energy difference $\Delta F$, providing knowledge of far-from-equilibrium thermodynamics.

The CFT has been verified in several classical systems~\cite{Schuler2005, Collin2005, Junier2009, Saira2012}, but a direct test in quantum systems remains elusive.
The difficulty originates from that work is not an observable in the quantum realm~\cite{Talkner2007}.
For isolated quantum systems, the work done during a process can be measured by the two-point measurement (TPM) protocol~\cite{Esposito2009, Campisi2011}.
The TPM protocol requires two high-fidelity non-demolition projective measurements (PMs) on the energy basis at the start and the end of the process to determine work.
Projective measurements with poor readout fidelity may not be able to obtain the initial and final energies correctly.
Measurements that are not non-demolition will result in the state after measurement different from the corresponding eigenstate.
Such situations will lead to incorrect work distribution and unable to recover the CFT~\cite{Hanggi2015}. 
However, experimental realization of high-fidelity non-demolition PM is generally challenging.
To evade the difficulty, some alternative but indirect approaches were proposed~\cite{Mazzola2013,Dorner2013} to obtain the work distribution and adopted by several experiments.
In a liquid-state nuclear magnetic resonance setup, the work distribution was reconstructed by a characteristic function which was measured using an auxiliary qubit~\cite{Batalh2014, Batalh2015}.
Besides, a pre-sampling method was utilized to test some integral fluctuation theorems in the nitrogen-vacancy (NV) center system\cite{Hernandez2020,Hernandez2021,Hernandez2022}.
To date, an experiment that faithfully implements the TPM protocol to test the CFT is still absent.

Here we report an experimental test of the CFT in a single nuclear spin, i.e., the ${}^{14}$N nuclear spin of the NV center in diamond. 
To implement the TPM protocol, high fidelity non-demolition projective measurements of the ${}^{14}$N nuclear spin were realized based on the single-shot readout technique~\cite{Neumann2010}.
The work statistics in the forward and time-reversed switching processes were experimentally obtained with the TPM protocol.
Our results demonstrate that the obtained work statistics satisfy the CFT for various speeds of the switching process and under different effective temperatures of the initial thermal state.

\begin{figure}
\centering
\includegraphics[width=0.9\textwidth]{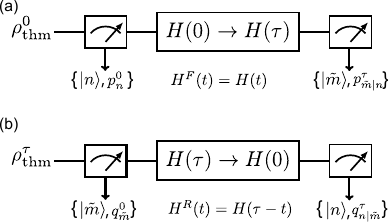}
\caption{The TPM protocol to obtain work distributions. 
(a) The forward switching process. $\rho^0_{\rm{thm}}$ is a thermal state of $H(0)$. The first and second projective measurements were performed at the energy eigenstates of $H(0)$ and $H(\tau)$, respectively. 
(b) The corresponding time-reversed switching process. $\rho^\tau_{\rm{thm}}$ is a thermal state of $H(\tau)$. The first and second projective measurements were performed at the energy eigenstates of $H(\tau)$ and $H(0)$, respectively.
}
\label{fig1}
\end{figure}

\section{II. Theory}
To test the CFT, work distributions of an isolated quantum system undergoing the forward and corresponding time-reversed switching processes are measured via the TPM protocol.
The procedure to obtain the work distribution in the forward switching process is shown in Fig.~\ref{fig1}(a).
Firstly, the system is prepared in the thermal state of $H(0)$, $\rho^0_{\rm{thm}}=e^{-\beta H(0)}/Z^0$, with $Z^0={\rm{Tr}}[e^{-\beta H(0)}]$ denoting the partition function.
Then, the first projective measurement is performed, projecting the system onto an energy eigenstate of $H(0)$, such as $|n\rangle$, with probability $p_n^{0}={\rm{Tr}}[\rho^0_{\rm{thm}}|n\rangle\langle n|]$.
Next, the system undergoes the forward switching process.
During this process, the Hamiltonian varies from $H(0)$ to $H(\tau)$ in a period of $\tau$.
The time-dependent Hamiltonian drives the system, for example, from $|n\rangle\langle n|$ into $\rho^\tau_n$.
Finally, the second projective measurement is performed, projecting the system onto an energy eigenstate of $H(\tau)$, such as $|\tilde{m}\rangle$, and the corresponding probability is $p_{\tilde{m}|n}^{\tau}={\rm{Tr}}[\rho^\tau_n|\tilde{m}\rangle\langle \tilde{m}|]$.
Then, the work done on the system for trajectory $|n\rangle \to|\tilde{m}\rangle$ is obtained as $W_{\tilde{m}|n} = E_{\tilde{m}}^\tau-E_n^0$, with $E_{\tilde{m}}^\tau$ and $E_n^0$ denoting the eigenenergies of states $|\tilde{m}\rangle$ and $|n\rangle$, respectively.
The work distribution in the forward switching process can be represented as $P^F(W) =\sum_{\tilde{m},n}p_n^{0}\cdot p_{\tilde{m}|n}^{\tau}\cdot\delta(W-W_{\tilde{m}|n})$.
The corresponding time-reversed switching process is displayed in Fig.~\ref{fig1}(b).
The system is initially prepared in the thermal state of $H(\tau)$, $\rho^{\tau}_{\rm{thm}}=e^{-\beta H(\tau)}/Z^\tau$ with $Z^\tau={\rm{Tr}}[e^{-\beta H(\tau)}]$.
The Hamiltonian is tuned from $H(\tau)$ to $H(0)$ in the time-reversed manner $H^R(t)=H(\tau-t)$.
The work distribution in the time-reversed switching process can be represented as $P^R(W) =\sum_{\tilde{m},n}q_{\tilde{m}}^{0}\cdot q_{n|\tilde{m}}^{\tau}\cdot\delta(W-W_{n|\tilde{m}})$.
The left-hand side of the CFT, $P^F(W)/P^R(-W)$, can then be calculated from the obtained work distributions in the forward and time-reversed processes.
The right-hand side of the CFT, $e^{\beta(W-\Delta F)}$, can be obtained via the Hamiltonian model $H(t)$ and the inverse temperature $\beta$.
The CFT is tested by checking whether the difference, $\Delta = P^F(W)/P^{R}(-W)-e^{\beta(W-\Delta F)}$, equals zero for all possible $W$. 

\section{III. Experiments}

\begin{figure*}
\centering
\includegraphics[width=0.8\textwidth]{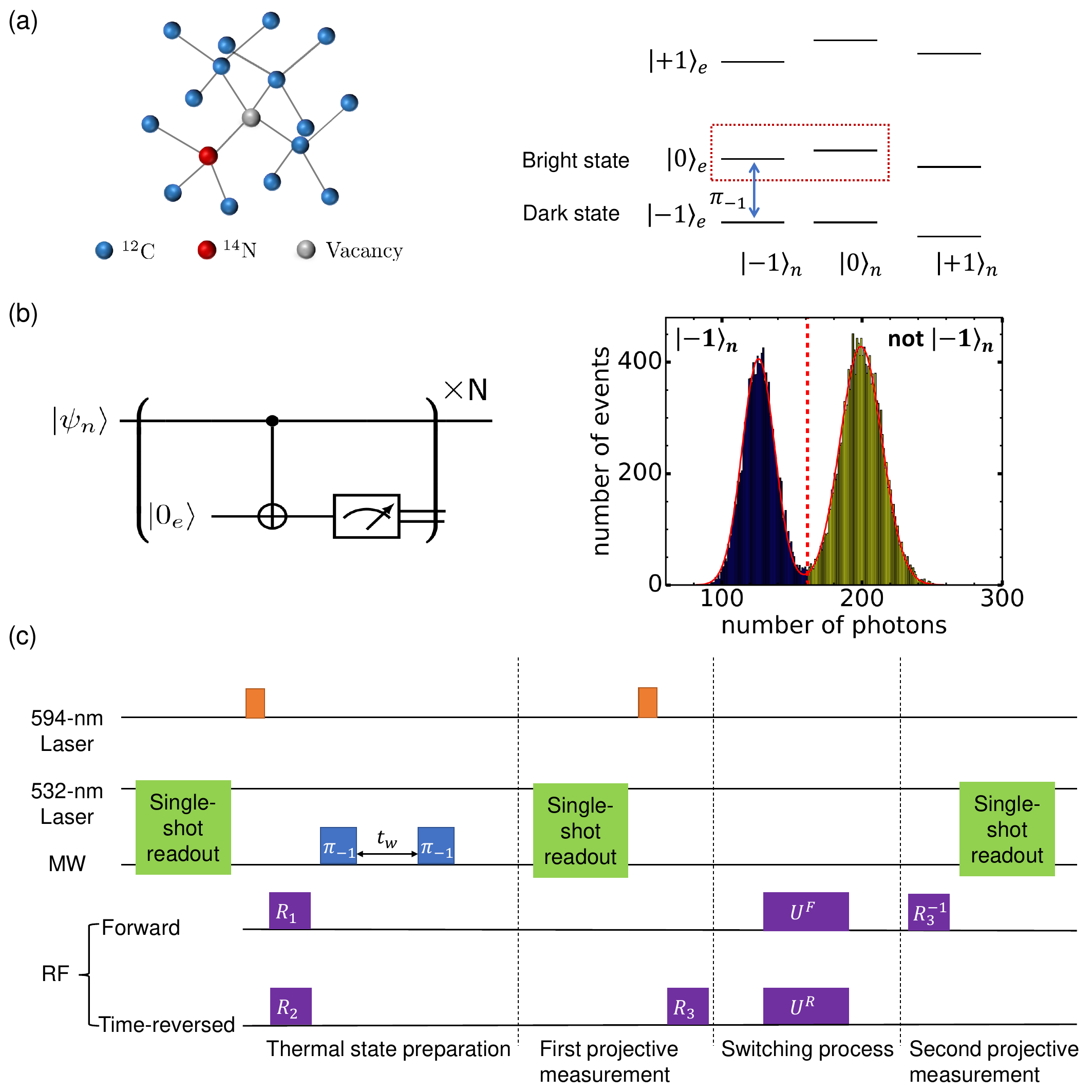}
\caption{Single-shot readout of the nuclear spin and realizing TPM in an NV center.
(a) Atomic structure and ground state energy levels of the NV center. 
The two energy levels in the dashed red box are utilized to test the CFT. 
(b) Single-shot readout.
Left: the pulse sequence of single-shot readout, here $\rm{N}=1500$. 
Right: the photon-counting histogram obtained by repeating single-shot readout. 
(c) Experimental pulse sequences to realize TPM to test the CFT.
}
\label{fig2} 
\end{figure*}

In our experiment, a single nuclear spin of the NV center was utilized to test the CFT.
The NV center is a type of defect in diamond consisting of a substitutional nitrogen atom adjacent to a carbon vacancy.
The left of Fig.~\ref{fig2}(a) shows the atomic structure and ground state energy levels of the NV center.
When a magnetic field is applied along the symmetry axis of the NV center, the ground state Hamiltonian can be written as
\begin{equation}
H_{\rm{NV}} = 2\pi\hbar(DS_{z}^{2}+\omega_{e}S_{z}+QI_{z}^{2}+\omega_{n}I_{z}+A_{zz}I_{z}S_{z})  , 
\end{equation}
where $S_{z}$ and $I_{z}$ are the spin operators of the $\rm{NV}$ electron spin and ${}^{14}$N nuclear spin, respectively. 
The ground-state zero-field splitting of electron spin is $D = 2.87$~GHz and the quadrupolar interaction of nuclear spin is $Q = -4.95$~MHz. 
The longitudinal hyperfine interaction between nuclear spin and electron spin is $A_{zz} = -2.16$~MHz. 
The Zeeman frequencies of the electron and nuclear spin induced by the external static magnetic field are denoted by $\omega_{e}$ and $\omega_{n}$, respectively.
The electron spin can be polarized into $|0\rangle_e$ via a spin-selective intersystem crossing process~\cite{Jacques2009}.
Due to the same mechanism, the photoluminescence rate for $|0\rangle_{e}$ is higher than that for $|-1\rangle$.
We denote $|0\rangle_{e}$ as the bright state and $|-1\rangle_{e}$ as the dark state in the following.
Two energy levels of the ${}^{14}$N nuclear spin, $|-1\rangle_{n}$ and $|0\rangle_{n}$, are chosen to form a two-level system to test the CFT as shown by the red dashed box in Fig.~\ref{fig2}(a).

High-fidelity non-demolition projective measurement of the nuclear spin was realized via the single-shot readout technique.
The single-shot readout process is displayed in Fig.~\ref{fig2}(b). 
The electron spin is optically pumped into the bright state.
Then, a selective $\pi_{-1}$ pulse flips the electron spin to the dark state conditioned that the nuclear spin state is $|-1\rangle_{n}$.
Next, a 532-nm laser pulse is applied to read out the electron spin and re-polarize it to the bright state.
By repeating this procedure, a fluorescence signal can be accumulated to read out the nuclear spin.
In the ideal case, the nuclear spin can be projectively measured. 
In practice, the projected state could be altered during the readout process due to the nuclear spin relaxation~\cite{Neumann2010}.
To suppress the relaxation, we applied a static magnetic field of approximately 7500 G along the NV symmetry axis.
Besides, the imperfection of the selective $\pi_{-1}$ pulse will also corrupt the fidelity.
Thus, a noise-robust gate was designed via an optimal control method.
The repetition number N was also appropriately chosen to optimize the fidelity.
With these techniques, the optimized fidelity achieved 0.98(1) (see Appendix A for details).
To realize projective measurements along arbitrary energy bases, necessary rotations can be applied before and after the single-shot readout.

The experimental pulse sequences are depicted in Fig.~\ref{fig2}(c).
By performing the single-shot readout and post-selecting the states with fluorescence below the threshold, the nuclear spin can be initialized into $|-1\rangle_{n}$.
During the readout process, the applied 532-nm laser pulse can induce transitions between two charge states $\rm{NV}^{0}$ and $\rm{NV}^{-}$ of the NV center.
A 594-nm laser pulse was applied to post-select the experiment trials done with $\rm{NV}^{-}$\cite{Aslam2013}.
In this work, the Hamiltonian of the switching process was chosen as
\begin{equation}
\left\{
\begin{array}{l}
\displaystyle H^{F}(t)= 2\pi\hbar[Z(t)S_z^{\prime}+X(t)S_x^{\prime}] \\
\displaystyle H^{R}(t)= 2\pi\hbar[Z(\tau-t)S_z^{\prime}+X(\tau-t)S_x^{\prime}],
\end{array}
\right.
\label{2}
\end{equation}
with $S_z^{\prime}=(|1\rangle\langle1|-|0\rangle\langle0|)/2$, $S_x^{\prime}=(|1\rangle\langle0|+|0\rangle\langle1|)/2$, $Z(t)=2$~kHz, and $X(t)=5[1-\cos({\pi t/\tau})]/2$~kHz.
For convenience, the energy levels $|0\rangle_{n}$ and $|-1\rangle_{n}$ are relabeled as $|1\rangle$ and $|0\rangle$, respectively, here and in the following. 
The thermal state of $H(0)$ was generated by two steps. 
In the first step, a resonant radio-frequency pulse $R_1$ was applied to prepare the state $\sqrt{P_{\rm{thm}}^0}|0\rangle + \sqrt{P_{\rm{thm}}^1}|1\rangle$, where $P_{\rm{thm}}$ is the thermal population.
In the second step, two selective $\pi_{-1}$ pulses separated by a waiting time $t_w =10$~$\upmu$s were applied to dissipate the coherence.
The coherence dissipated quickly as the dephasing time of the electron spin $T_{2,e}^* < 1.5$~$\upmu$s.
After the preparation of the thermal state, the first projective measurements were performed to project the system onto energy eigenstates.
In the time-reversed process, the eigenstates of $H(0)$ differ from the computational basis, so an additional rotation pulse $R_3$ is applied after the single-shot readout.
Then, the time-dependent Hamiltonian or its time-reversed counterpart was applied to change the system state.
Some work was extracted from or performed on the system during the switching process.
Finally, the second projective measurements were performed to obtain the work trajectories. 
In the forward process, the eigenstates of $H(\tau)$ differ from the computational basis, so an additional rotation pulse $R_3^{-1}$ is applied before the single-shot readout.
To obtain the work statistics, the pulse sequence was executed 16,000 times for each experimental data point.
The post-selection ratio for the nuclear state initialization and charge state selection were 32\% and 41\%, respectively.
Since the charge state post-selection was executed twice in our experimental sequence, the total success ratio was about 5.4\%.
By analyzing the post-selected data, we can obtain the work statistics of these switching processes.

\begin{figure}
\centering
\includegraphics[width=.95\textwidth]{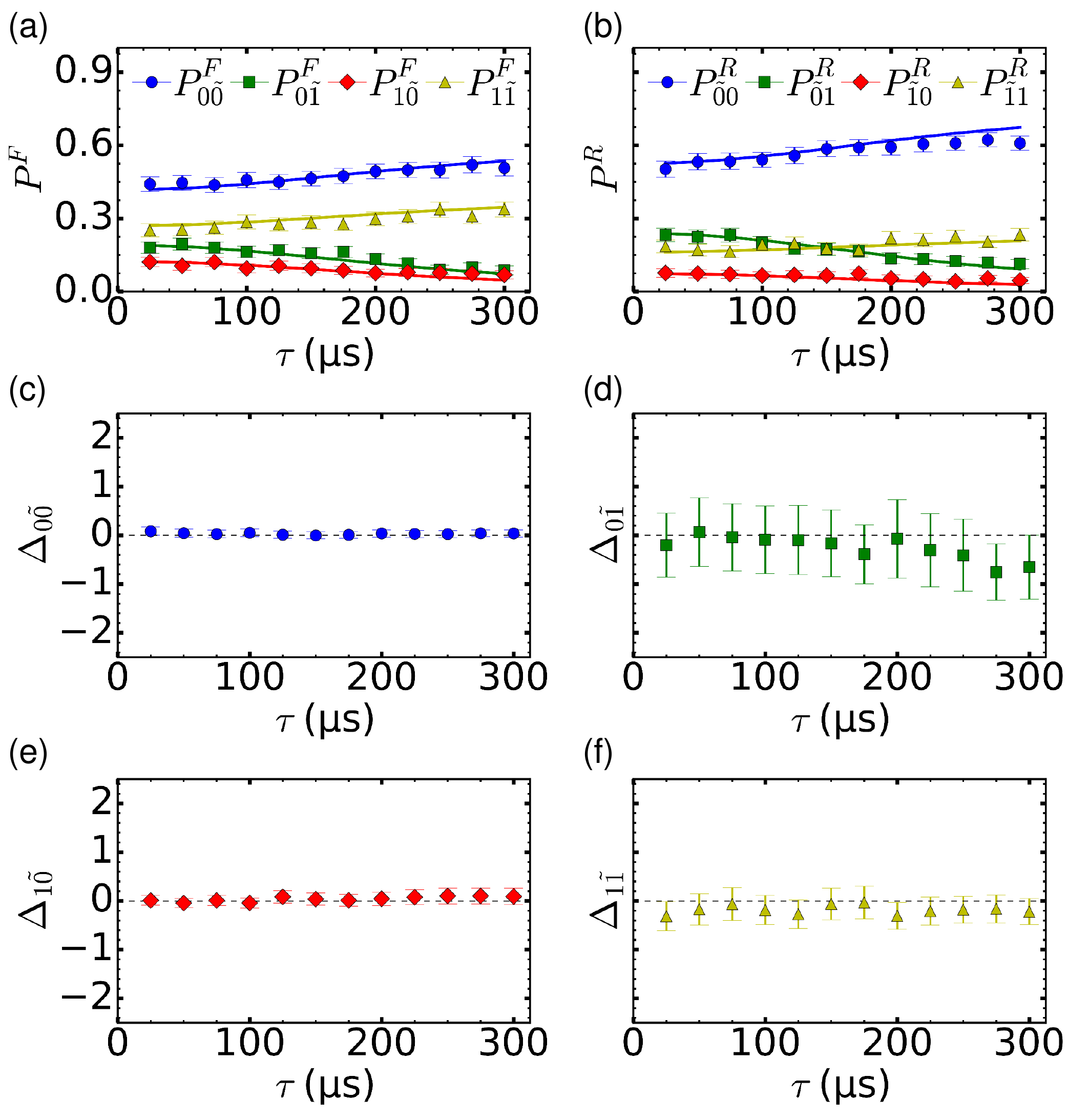}
\caption{Experimental verification of the CFT for different switching times.
(a)$\sim$(b) Probability distributions of trajectories in (a) the forward processes and (b) corresponding time-reversed processes for different switching times.
(c)$\sim$(f) Test of the CFT utilizing different trajectories, dots with error bars show the $\Delta_{i\tilde{j}} = P_{i\tilde{j}}^F/P_{\tilde{j}i}^{R}-e^{\beta(W_{i\tilde{j}}-\Delta F)}$ for different switching times.
}
\label{fig3}
\end{figure}

To test the CFT in switching processes with different adiabaticities, we conducted the experimental investigations with switching time $\tau$ ranging from 25~$\upmu$s to 300~$\upmu$s. 
The adiabaticity can be evaluated by parameter~\cite{Sakurai2014} $\Gamma=\min_{t\in[0,\tau]}|\langle n_1(t)|\partial H(t)/\partial t|n_2(t)\rangle|/[E_1(t)-E_2(t)]^2$.
Here $|n_1(t)\rangle$, $|n_2(t)\rangle$ are two instantaneous eigenstates of $H(t)$ with $E_1(t)$, $E_2(t)$ being the corresponding energies. 
When $\Gamma$ is much smaller than 1, the switching process can be considered adiabatic.
In our experiment, the switching process varies from a fast process to a close-to-adiabatic process with $\Gamma$ decreasing from 3.6 to 0.3.
The experimental probability distributions of trajectories in the forward process and corresponding time-reversed process are displayed in Fig.~\ref{fig3}(a) and \ref{fig3}(b), respectively. 
$P_{i\tilde{j}}^F$ refers to the probability of trajectory $|i\rangle \to|\tilde{j}\rangle$ in the forward process and $P_{\tilde{j}i}^R$ refers to the probability of trajectory $|\tilde{j}\rangle \to|i\rangle$ in the time-reversed process.
The error bars shown here represent the 95${\%}$ confidence interval.
As $\tau$ increases, the transition probabilities between different instantaneous eigenstates decrease, indicating that the process is approaching the adiabatic regime gradually.
To calculate the differences, $\Delta_{i\tilde{j}} = P_{i\tilde{j}}^F/P_{\tilde{j}i}^{R}-e^{\beta(W_{i\tilde{j}}-\Delta F)}$, the effective inverse temperature $\beta_{\rm{expt}}$ and free-energy difference $\Delta F$ were obtained from the measured initial thermal populations (see Appendix B for details).
Experimentally, the same initial thermal state was prepared for each switching time.
However, mainly due to the single-shot noises, the measured populations had non-zero uncertainties, leading to uncertainty in the value of $\beta_{\rm{expt}}$.
Here, the effective inverse temperature was $h\beta_{\rm{expt}} = 0.22(3)~({\rm{kHz}}^{-1})$.
The uncertainties of $\Delta_{i\tilde{j}}$ were obtained utilizing the error transfer formula.
As displayed in Fig.~\ref{fig3}(d), the uncertainties of $\Delta_{1\tilde{0}}$ are considerably larger than those of the other three, mainly because the denominator, $P_{\tilde{1}0}^{R}$, in $\Delta_{1\tilde{0}}$ is considerably smaller.
As shown by Fig.~\ref{fig3}(c)$\sim$(f), these experimental points can be considered equal to zero when error bars are taken into account, confirming the validity of the CFT for different speeds of the switching process.

\begin{figure}
\centering
\includegraphics[width=.95\textwidth]{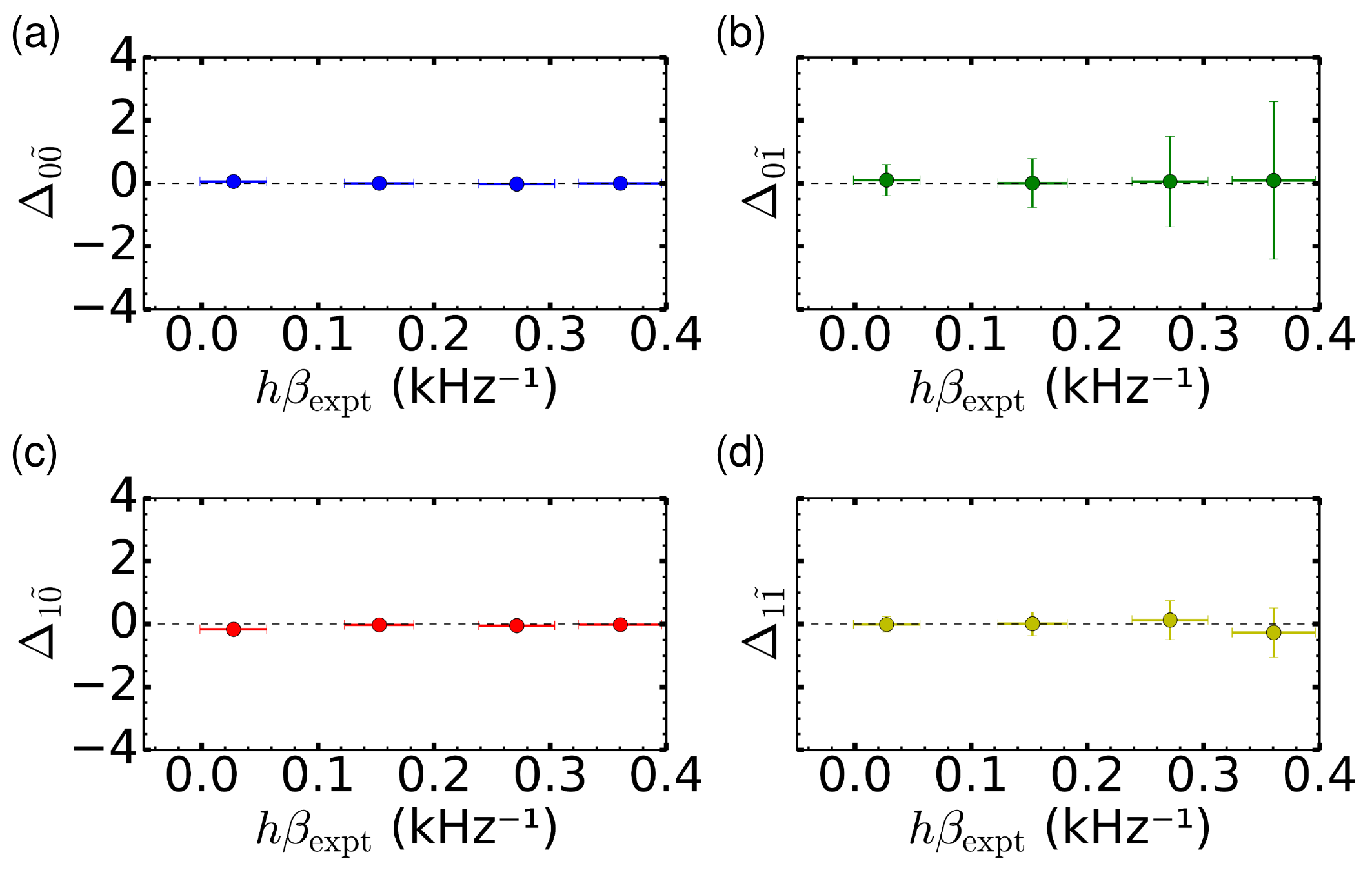}
\caption{Experimental verification of the CFT under different temperatures.
(a)$\sim$(d) Test of the CFT utilizing different trajectories, dots with vertical error bars show the $\Delta_{i\tilde{j}} = P_{i\tilde{j}}^F/P_{\tilde{j}i}^{R}-e^{\beta(W_{i\tilde{j}}-\Delta F)}$ under different effective temperatures. 
The horizontal error bars represent the errors of $h\beta_{\rm{expt}}$.
}
\label{fig4}
\end{figure}

Furthermore, to test the CFT under different temperatures, we prepared different initial thermal states.
Our experiment fixed the switching time at $\tau = 25$~$\upmu$s.
The inverse temperatures were preset as $h\beta= 0,0.15,0.25,0.35~({\rm{kHz}}^{-1})$, with $h\beta=0~({\rm{kHz}}^{-1})$ representing an infinitely high temperature.
The effective inverse temperature were $h\beta_{\rm{expt}}=0.03(3), 0.15(3), 0.27(3), 0.36(4)~({\rm{kHz}}^{-1})$.
For different effective temperatures, the difference $\Delta_{i\tilde{j}}$ and their uncertainties were calculated.
As shown by Fig.~\ref{fig4}(a)$\sim$(d), these experimental points can be considered as equal to zero when error bars are taken into account, verifying the CFT under different temperatures.

\section{IV. Conclusion}
In conclusion, we experimentally tested the CFT by faithfully implementing the TPM protocol.
The work distributions in the forward and corresponding time-reversed switching processes were measured and the difference between the left- and right-hand sides of the CFT was obtained. 
The experimental results show that the difference is zero for different speeds of nonequilibrium processes and under various effective temperatures, providing a rigorous validation of the CFT.
Additionally, our development of high-fidelity non-demolition projective measurement in the NV center system can facilitate the investigation of quantum thermodynamics, enabling us to study many other important thermodynamic principles and interesting phenomena.
For example, the fluctuation theorem in the non-Hermitian regime~\cite{Zeng2017}, information-theoretic-based quantum thermodynamics~\cite{Kim2011, Goold2016}, and quantum thermodynamic devices~\cite{Kosloff2014, Binder2015, Myers2022} can be further explored.
It is noted that the TPM protocol has been regarded to destroy quantum features~\cite{Xu2018}.
Recently in bipartite systems, fluctuations of heat exchange were studied beyond the framework of the TPM protocol to consider the influence of quantum correlation~\cite{Micadei2020,Micadei2021}.
There have also been other attempts to explore alternative measurement protocols~\cite{Roncaglia2014,Allahverdyan2014,Solinas2017} that account for initially coherent states.
However, in our test of the quantum CFT, the system needs to be prepared in a thermal state with no quantum coherence or quantum correlation, and therefore the TPM protocol is an appropriate and standard method.

\section{ACKNOWLEDGMENTS}
We thank Yang Wu for the helpful discussions. This work was supported by the National Key R\&D Program of China (Grants No. 2018YFA0306600 and Grants No. 2016YFB0501603), the National Natural Science Foundation of China (Grants No. 12174373), the Chinese Academy of Sciences (Grants No. XDC07000000 and Grants No. GJJSTD20200001), Innovation Program for Quantum Science and Technology (Grants No. 2021ZD0302200), Anhui Initiative in Quantum Information Technologies (Grants No. AHY050000) and Hefei Comprehensive National Science.
Center. X.R. thanks the Youth Innovation Promotion Association of Chinese Academy of Sciences for their support. W.L. is funded by Beijing University of Posts and Telecommunications Innovation Group.

W.C. and W.L. contributed equally to this work.

\section{APPENDIX A: High fidelity projective measurement}
To improve the fidelity, the measurement back-action should be mitigated and the detection efficiency should be improved.
We applied a static magnetic field of about 7500 G along the NV symmetry axis to mitigate the measurement back-action.
We created a solid immersion lens~\cite{Hadden2010} in the diamond and used an oil objective to improve the detection efficiency.
Furthermore, the optimal control method~\cite{Rembold2020} was utilized to realize a noise-robust gate in the single-shot readout. 

\begin{figure}
\centering
\includegraphics[width=.9\textwidth]{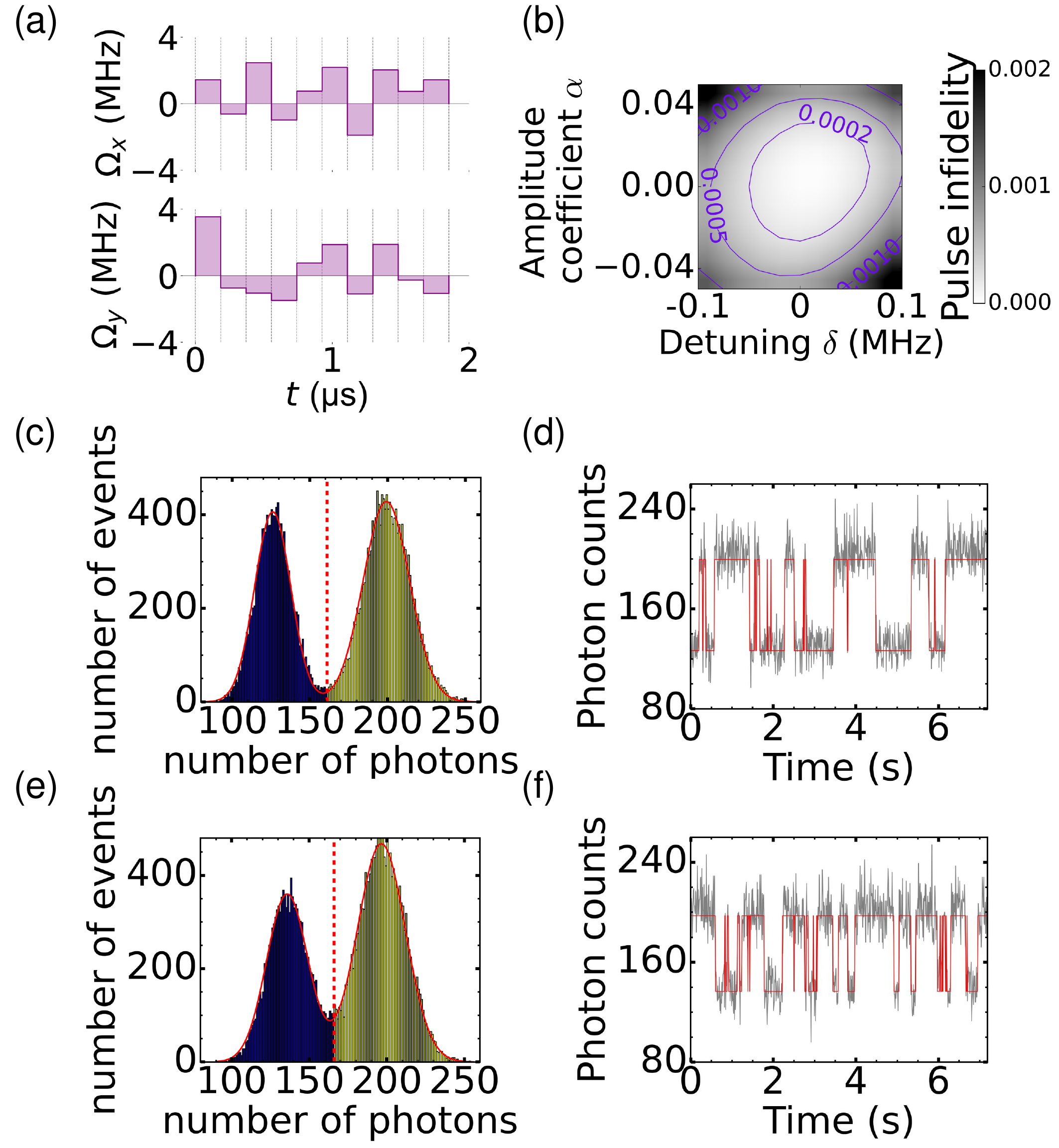}
\caption{Single-shot readout with optimal control.
(a)$\sim$(b) Optimized pulse and its robustness to noise. 
(c)$\sim$(f) Photon-counting histogram and fluorescence time trace with an optimized pulse (c)$\sim$(d) and with a naive square-wave pulse (e)$\sim$(f).
}
\label{fig5}
\end{figure}

We studied the optimal quantum control in the Hilbert space spanned by $|1\rangle_n|0\rangle_e$, $|0\rangle_n|0\rangle_e$, $|-1\rangle_n|0\rangle_e$, $|1\rangle_n|-1\rangle_e$, $|0\rangle_n|-1\rangle_e$ and $|-1\rangle_n|-1\rangle_e$.  
Microwave pulses, whose frequency equals the energy difference between $|0\rangle_n|0\rangle_e$ and $|-1\rangle_n|0\rangle_e$, were applied to control the NV electron spin.
In the rotational frame, the system Hamiltonian is $H_0=2\pi A_{zz}I_z\otimes|-1\rangle_e\langle\ -1|$. 
The control Hamiltonian is a piecewise constant. 
Denoting $t_i = t_0 + i\Delta t$, the control Hamiltonian at $t\in[t_{i-1},t_i)$ takes the form
\begin{align}
H_c(t) = 2\pi[\Omega_x(t)H_x+\Omega_y(t)H_y]=2\pi(\Omega_{ix}H_x+\Omega_{iy}H_y),
\end{align}
where $\Omega_{ix}$ and  $\Omega_{iy}$ are parameters to be optimized, and $H_{x,y}=\mathbf{1}_n\otimes S^{\prime}_{x,y}$.
The amplitude noise of the control field leads to $H_c^{\prime}(t)=(1+\alpha)H_c(t)$ and the static dephasing noise takes the form $H_d=\mathbf{1}_n\otimes 2\pi\delta S_z^{\prime}$.
The total Hamiltonian is $H(t)=H_0+(1+\alpha)H_c(t)+H_d$.
The evolution time is set as $T=4/|A_{zz}|$ and is divided into $M=10$ segments of equal length. 
The propagator of these pulses is $U(T)=\Pi_{i}e^{-jH_iT/M}$, where $H_i=H_0+(1+\alpha)\cdot2\pi(\Omega_{ix}H_x+\Omega_{iy}H_y)+H_d$.
The target gate is $U_{targ} = |-1\rangle_n\langle\ -1|\otimes e^{-j\pi S_x^{\prime}}+(|1\rangle_n\langle1|+|0\rangle_n\langle0|)\otimes\mathbf{1}_e$.
The fidelity between $U$ and $U_{targ}$ is defined as $\mathcal{F}=\big|{\rm{Tr}}[U_{targ}^{\dagger}U]/{\rm{Tr}}[U^{\dagger}U]\big|^2$.
We designed a noise-robust pulse that can realize a high fidelity $U$ with arbitrary $\alpha$ and $\delta$, where $\alpha\in[\alpha_{min},\alpha_{max}]$ and $\delta\in[\delta_{min},\delta_{max}]$.
Pulse shape and its robustness are shown in Fig.~\ref{fig5}(a) and Fig.~\ref{fig5}(b), respectively.

Photon-counting histograms and fluorescence time traces with both optimized pulse and naive square-wave pulse are shown in Fig.~\ref{fig5}(c)$\sim$(f).
In the fluorescence time trace, each data point was acquired by the single-shot readout with repetition number $N=1500$ (total time 9\,$\rm{ms}$).
There is a telegraph-like signal in the fluorescence time trace, and two plateaus represent that the nuclear spin stays in or not in $|-1\rangle_n$.
The transition between the two plateaus comes from either misjudgment or intrinsic flips of the nuclear spin.
The average data point of the nuclear spin being in $|-1\rangle_n$ is $\bar{n}_0$ and the nuclear spin being not in $|-1\rangle_n$ is $\bar{n}_1$.
The fidelity can be defined as $(F_{0,1})^2=1-1/\bar{n}_{0,1}$.
In our experiment $\bar{n}_{0,1}>>1$ and the average fidelity is approximately
\begin{equation}
F = 1 - \frac{1}{2}(\frac{1}{2\bar{n}_{0}}+\frac{1}{2\bar{n}_{1}}),
\end{equation}
Fidelities with optimized pulse and naive square wave pulse are $F_{opt}=0.98(1)$ and $F_{squ}=0.96(1)$, respectively.

\section{APPENDIX B: Calculation of $\beta_{\rm{expt}}$ and $\Delta F$}
Utilizing a single 2-level system, four potential work trajectories exist during a switching process in our experiment.
The work distribution and its uncertainty were obtained by counting these four trajectories.
The effective inverse temperature $\beta_{\rm{expt}}$ and its error were calculated by the work distribution and its uncertainty. 
For convenience, we denote $P^F(W=E_{\tilde{j}}^{\tau}-E_i^0)$ as $P_{i\tilde{j}}$ in the following.
To obtain $\beta_{\rm{expt}}$, we calculated the initial population $p_i$ by summing $P_{i\tilde{j}}$ over index $j$, $p_i = \sum_jP_{i\tilde{j}}$.
In the forward process, $\beta_{\rm{expt}}^{F}$ is given by $\beta_{\rm{expt}}^{F}={\rm{In}}(p_0/p_1)/(E_1^0-E_0^0)$, and its error can be calculated using error transfer formula.
In the time-reversed process, $\beta_{\rm{expt}}^{R}$ and its error were obtained similarly.
The average effective inverse temperature is $\beta_{\rm{expt}}=(\beta_{\rm{expt}}^{F}+\beta_{\rm{expt}}^{R})/2$.
Upon obtaining $\beta_{\rm{expt}}$, the free-energy difference can be calculated as $\Delta F=-{\rm{In}}[{\rm{Tr}}[e^{-\beta_{\rm{expt}}H(\tau)}]/{\rm{Tr}}[e^{-\beta_{\rm{expt}}H(0)}]]/\beta_{\rm{expt}}$.
In the test of the CFT for different speeds of switching processes, the effective inverse temperature was $h\beta_{\rm{expt}} = 0.22(3)~({\rm{kHz}}^{-1})$, with $h\beta_{\rm{expt}}^{F} = 0.24(5)~({\rm{kHz}}^{-1})$ and $h\beta_{\rm{expt}}^{R} = 0.20(2)~({\rm{kHz}}^{-1})$.
In the test of the CFT under different temperatures, $h\beta_{\rm{expt}}$ and corresponding initial populations are listed in Table.\ref{table}.
The errosr of $h\beta^R_{\rm{expt}}$ are smaller than $h\beta^F_{\rm{expt}}$ due to the larger energy gap of $H(\tau)$ than $H(0)$.

\begin{table}
\renewcommand{\tablename}{Table}
\begin{tabular}{ccccccc}
\hline
$h\beta_{\rm{expt}}$ & $p_0$ & $p_1$ & $h\beta^F_{\rm{expt}}$ & $q_0$ & $q_1$ & $h\beta^R_{\rm{expt}}$\\
\hline
0.03(3) & 0.52(4) & 0.48(4) & 0.04(4) & 0.53(4) & 0.47(4) & 0.02(2) \\
0.15(3) & 0.58(4) & 0.42(4) & 0.16(5) & 0.69(4) & 0.31(4) & 0.15(2) \\
0.27(3) & 0.63(4) & 0.37(4) & 0.27(6) & 0.81(3) & 0.19(3) & 0.27(3) \\
0.36(4) & 0.68(4) & 0.32(4) & 0.38(6) & 0.86(3) & 0.14(3) & 0.34(4) \\
\hline
\end{tabular}
\caption{\label{table}
\raggedright
Effective inverse temperatures and corresponding initial populations of forward and time-reversed process.
$h\beta^F_{\rm{expt}}$ and $h\beta^R_{\rm{expt}}$ were obtained from initial populations. $h\beta_{\rm{expt}}$ were obtained as $h\beta_{\rm{expt}}=(h\beta^F_{\rm{expt}}+h\beta^R_{\rm{expt}})/2$
}
\end{table}


\begin{thebibliography}{99}

\bibitem{Baumer2018} E. B{\"a}umer, M. Lostaglio, M. Perarnau-Llobet, and R. Sampai, Fluctuating work in coherent quantum systems: Proposals and limitations, in \href{https://link.springer.com/chapter/10.1007/978-3-319-99046-0_11}{Thermodynamics in the Quantum Regime} (Springer,2018) pp.275--300

\bibitem{Jarzynski2011} C. Jarzynski, Equalities and inequalities: Irreversibility and the second law of thermodynamics at the nanoscale, \href{\doibase 10.1146/annurev-conmatphys-062910-140506}{Annual Review of Condensed Matter Physics \textbf{2}, 603--633 (2011).}

\bibitem{Sevick2008} E. Sevick, R. Prabhakar, S. R. Williams, and D. J. Searles, Fluctuation theorems, \href {\doibase 10.1146/annurev.physchem.58.032806.104555}{Annual Review of Physical Chemistry \textbf{59}, 329--351 (2008).}

\bibitem{Esposito2009} M. Esposito, U. Harbola, and S. Mukamel, Nonequilibrium fluctuations, fluctuation theorems, and counting statistics in quantum systems, \href{https://link.aps.org/doi/10.1103/RevModPhys.81.1665}{Rev. Mod. Phys. \textbf{81}, 1665--1702 (2009).}

\bibitem{Campisi2011} M. Campisi, P. H{\"a}nggi, and P. Talkner, Colloquium: Quantum fluctuation relations: Foundations and applications, \href{https://link.aps.org/doi/10.1103/RevModPhys.83.771}{Rev. Mod. Phys. \textbf{83}, 771 (2011).}

\bibitem{Funo2018} K. Funo, M. Ueda, and T. Sagawa, Quantum Fluctuation Theorems, in \href{https://link.springer.com/chapter/10.1007/978-3-319-99046-0_10}{Thermodynamics in the Quantum Regime} (Springer,2018) pp.249--273

\bibitem{Landi2021} G. T. Landi and M. Paternostro, Irreversible entropy production: From classical to quantum, \href{https://link.aps.org/doi/10.1103/RevModPhys.93.035008}{Rev. Mod. Phys. \textbf{93}, 035008 (2021).}

\bibitem{Crooks1999} G. E. Crooks, Entropy production fluctuation theorem and the nonequilibrium work relation for free energy differences, \href{https://link.aps.org/doi/10.1103/PhysRevE.60.2721}{Phys. Rev. E \textbf{60}, 2721--2726 (1999).}

\bibitem{Schuler2005} S. Schuler, T. Speck, C. Tietz, J. Wrachtrup, and U. Seifert, Experimental test of the fluctuation theorem for a driven two-level system with time-dependent rates, \href{https://link.aps.org/doi/10.1103/PhysRevLett.94.180602}{Phys. Rev. Lett. \textbf{94}, 180602 (2005).}

\bibitem{Collin2005} D. Collin, F. Ritort, C. Jarzynski, S. B. Smith, I. Tinoco, and C. Bustamante, Verification of the Crooks fluctuation theorem and recovery of RNA folding free energies, \href{https://www.nature.com/articles/nature04061}{Nature \textbf{437}, 231--234 (2005).}

\bibitem{Junier2009} I. Junier, A. Mossa, M. Manosas, and F. Ritort, Recovery of free energy branches in single molecule experiments, \href{https://link.aps.org/doi/10.1103/PhysRevLett.102.070602}{Phys. Rev. Lett. \textbf{102}, 070602 (2009).}

\bibitem{Saira2012} O.-P. Saira, Y. Yoon, T. Tanttu, M. M\"ott\"onen, D. V. Averin, and J. P. Pekola, Test of the Jarzynski and Crooks fluctuation relations in an electronic system, \href{https://link.aps.org/doi/10.1103/PhysRevLett.109.180601}{Phys. Rev. Lett. \textbf{109}, 180601 (2012).}

\bibitem{Talkner2007} P. Talkner, E. Lutz, and P. H\"anggi, Fluctuation theorems: Work is not an observable, \href{https://link.aps.org/doi/10.1103/PhysRevE.75.050102}{Phys. Rev. E \textbf{75}, 050102(R) (2007).}

\bibitem{Hanggi2015} P. H{\"a}nggi, and P. Talkner, The other QFT, \href{https://www.nature.com/articles/nphys3167}{Nature Physics \textbf{11}, 108--110 (2015).}

\bibitem{Mazzola2013} L. Mazzola, G. De Chiara, and M. Paternostro, D. V. Averin, and J. P. Pekola, Measuring the characteristic function of the work distribution, \href{https://link.aps.org/doi/10.1103/PhysRevLett.110.230602}{Phys. Rev. Lett. \textbf{110}, 230602 (2013).}

\bibitem{Dorner2013} R. Dorner, S. R. Clark, L. Heaney, R. Fazio, J. Goold, and V. Vedral, Extracting quantum work statistics and fluctuation theorems by single-qubit interferometry, \href{https://link.aps.org/doi/10.1103/PhysRevLett.110.230601}{Phys. Rev. Lett. \textbf{110}, 230601 (2013).}

\bibitem{Batalh2014} T. B. Batalh\~ao, A. M. Souza, L. Mazzola, R. Auccaise, R. S. Sarthour, I. S. Oliveira, J. Goold, G. De Chiara, M. Paternostro, and R. M. Serra, Experimental reconstruction of work distribution and study of fluctuation relations in a closed quantum system, \href{https://link.aps.org/doi/10.1103/PhysRevLett.113.140601}{Phys. Rev. Lett. \textbf{113}, 140601 (2014).}

\bibitem{Batalh2015} T. B. Batalh\~ao, A. M. Souza, R. S. Sarthour, I. S. Oliveira, M. Paternostro, E. Lutz, and R. M. Serra, Irreversibility and the arrow of time in a quenched quantum system, \href{https://link.aps.org/doi/10.1103/PhysRevLett.115.190601}{Phys. Rev. Lett. \textbf{115}, 190601 (2015).}

\bibitem{An2015} S. An, J.-N. Zhang, M. Um, D. Lv, Y. Lu, J. Zhang, Z.-Q. Yin, H. Quan, and K. Kim, Experimental test of the quantum Jarzynski equality with a trapped-ion system, \href{https://www.nature.com/articles/nphys3197}{Nature Physics \textbf{11}, 193--199 (2015).}

\bibitem{Smith2018} A. Smith, Y. Lu, S. An, X. Zhang, J.-N. Zhang, Z. Gong, H. T. Quan, C. Jarzynski, and K. Kim, Verification of the quantum nonequilibrium work relation in the presence of decoherence, \href{https://dx.doi.org/10.1088/1367-2630/aa9cd6}{New Journal of Physics \textbf{20}, 013008 (2018).}

\bibitem{Hernandez2020} S. Hern\'andez-G\'omez, A. M. Souza, R. S. Sarthour, I. S. Oliveira, M. Paternostro, E. Lutz, and R. M. Serra, Experimental test of exchange fluctuation relations in an open quantum system, \href{https://link.aps.org/doi/10.1103/PhysRevResearch.2.023327}{Phys. Rev. Res. \textbf{2}, 023327 (2020).}

\bibitem{Hernandez2021} S. Hern\'andez-G\'omez, N. Staudenmaier, M. Campisi, and N. Fabbri, Experimental test of fluctuation relations for driven open quantum systems with an NV center, \href{https://dx.doi.org/10.1088/1367-2630/abfc6a}{New Journal of Physics \textbf{23}, 065004 (2021).}

\bibitem{Hernandez2022} S. Hern\'andez-G\'omez, S. Gherardini, N. Staudenmaier, F. Poggiali, M. Campisi, A. Trombettoni, F. Cataliotti, P. Cappellaro, and N. Fabbri, Autonomous dissipative Maxwell’s demon in a diamond spin qutrit, \href{https://link.aps.org/doi/10.1103/PRXQuantum.3.020329}{PRX Quantum \textbf{3}, 020329 (2022).}

\bibitem{Neumann2010} P. Neumann, J. Beck, M. Steiner, F. Rempp, H. Fedder, P. R. Hemmer, J. Wrachtrup, and F. Jelezko, Single-shot Readout of a Single Nuclear Spin, \href{https://www.science.org/doi/10.1126/science.1189075}{Science \textbf{329}, 542--544 (2010).}

\bibitem{Jacques2009} V. Jacques, P. Neumann, J. Beck, M. Markham, D. Twitchen, J. Meijer, F. Kaiser, G. Balasubramanian, F. Jelezko, and J. Wrachtrup, Dynamic polarization of single nuclear spins by optical pumping of nitrogen-vacancy color centers in diamond at room temperature, \href{https://link.aps.org/doi/10.1103/PhysRevLett.102.057403}{Phys. Rev. Lett. \textbf{102}, 057403 (2009).}

\bibitem{Aslam2013} N. Aslam, G. Waldherr, P. Neumann, F. Jelezko, and J. Wrachtrup, Photo-induced ionization dynamics of the nitrogen vacancy defect in diamond investigated by single-shot charge state detection, \href{https://dx.doi.org/10.1088/1367-2630/15/1/013064}{New Journal of Physics \textbf{15}, 013064 (2013).}

\bibitem{Sakurai2014} J. Sakurai and J. Napolitano, \emph{Modern Quantum Mechanics} (Pearson Education, London, 2014).

\bibitem{Zeng2017} M. Zeng and E. H. Yong, Crooks fluctuation theorem in $\mathcal{P}$$\mathcal{T}$-symmetric quantum mechanics, \href{https://dx.doi.org/10.1088/2399-6528/aa8f26}{Journal of Physics Communications \textbf{1}, 031001 (2017).}

\bibitem{Kim2011} S. W. Kim, T. Sagawa, S. De Liberato, and M. Ueda, Quantum Szilard engine, \href{https://link.aps.org/doi/10.1103/PhysRevLett.106.070401}{Phys. Rev. Lett. \textbf{106}, 070401 (2011).}

\bibitem{Goold2016} J. Goold, M. Huber, A. Riera, L. del Rio, and P. Skrzypczyk, The role of quantum information in thermodynamics—a topical review, \href{https://dx.doi.org/10.1088/1751-8113/49/14/143001}{Journal of Physics A: Mathematical and Theoretical \textbf{49}, 143001 (2016).}

\bibitem{Kosloff2014} R. Kosloff and A. Levy, Quantum heat engines and refrigerators: Continuous devices, \href{\doibase 10.1146/annurev-physchem-040513-103724}{Annual Review of Physical Chemistry \textbf{65}, 365--393 (2014).}

\bibitem{Binder2015} F. C. Binder, S. Vinjanampathy, K. Modi, and J. Goold, Quantacell: powerful charging of quantum batteries, \href{https://dx.doi.org/10.1088/1367-2630/17/7/075015}{New Journal of Physics \textbf{17}, 075015 (2015).}

\bibitem{Myers2022} N. M. Myers, O. Abah, and S. Deffner, Quantum thermodynamic devices: From theoretical proposals to experimental reality, \href{https://doi.org/10.1116/5.0083192}{AVS Quantum Science \textbf{4}, 027101 (2022).}

\bibitem{Xu2018} B.-M. Xu, J. Zou, L.-S. Guo, and X.-M. Kong, Effects of quantum coherence on work statistics, \href{\doibase 10.1103/PhysRevA.97.052122}{Phys. Rev. A \textbf{97}, 052122 (2018).}

\bibitem{Micadei2020} K. Micadei, G. T. Landi, and E. Lutz, Quantum fluctuation theorems beyond two-point measurements, \href{https://link.aps.org/doi/10.1103/PhysRevLett.124.090602}{Phys. Rev. Lett. \textbf{124}, 090602 (2020).}

\bibitem{Micadei2021} K. Micadei, J. P. S. Peterson, A. M. Souza, R. S. Sarthour, I. S. Oliveira, G. T. Landi, R. M. Serra, and E. Lutz, Experimental validation of fully quantum fluctuation theorems using dynamic bayesian networks, \href{https://link.aps.org/doi/10.1103/PhysRevLett.127.180603}{Phys. Rev. Lett. \textbf{127}, 180603 (2021).}

\bibitem{Roncaglia2014} A. J. Roncaglia, F. Cerisola, and J. P. Paz, Work measurement as a generalized quantum measurement, \href{https://link.aps.org/doi/10.1103/PhysRevLett.113.250601}{Phys. Rev. Lett. \textbf{113}, 250601 (2014).}

\bibitem{Allahverdyan2014} A. E. Allahverdyan, Nonequilibrium quantum fluctuations of work, \href{https://link.aps.org/doi/10.1103/PhysRevE.90.032137}{Phys. Rev. E \textbf{90}, 032137 (2014).}

\bibitem{Solinas2017} P. Solinas, H. J. D. Miller, and J. Anders, Measurement-dependent corrections to work distributions arising from quantum coherences, \href{https://link.aps.org/doi/10.1103/PhysRevA.96.052115}{Phys. Rev. A \textbf{96}, 052115 (2017).}

\bibitem{Hadden2010} J. P. Hadden, J. P. Harrison, A. C. Stanley-Clarke, L. Marseglia, Y.-L. D. Ho, B. R. Patton, J. L. O’Brien, and J. G. Rarity, Strongly enhanced photon collection from diamond defect centers under microfabricated integrated solid immersion lenses, \href{https://doi.org/10.1063/1.3519847}{Applied Physics Letters \textbf{97}, 241901 (2010).}

\bibitem{Rembold2020} P. Rembold, N. Oshnik, M. M. M{\"u}ller, S. Montangero, T. Calarco, and E. Neu, Introduction to quantum optimal control for quantum sensing with nitrogen-vacancy centers in diamond, \href{https://doi.org/10.1116/5.0006785}{AVS Quantum Science \textbf{2}, 024701 (2020).}

\end{thebibliography}
\end{document}